\documentclass[journal]{IEEEtran}
\IEEEoverridecommandlockouts
\usepackage{stfloats}
\usepackage{cite}
\usepackage{version}
\usepackage{amsmath,amssymb,amsfonts}
\usepackage{amssymb,amsbsy,epsfig,float,bm}
\usepackage{algorithmic}
\usepackage{graphicx}
\usepackage{textcomp}
\usepackage{xcolor}

\def\BibTeX{{\rm B\kern-.05em{\sc i\kern-.025em b}\kern-.08em
		T\kern-.1667em\lower.7ex\hbox{E}\kern-.125emX}}
\usepackage{wrapfig}
\newtheorem{thm}{Theorem}

\newtheorem{proposition}{Proposition}
\newtheorem{cor}{Corollary}

\usepackage[caption=false,font=footnotesize]{subfig}

\begin{document}
	
	\title{Profit Sharing Contracts between Content and Service Providers for Enhanced Network Quality }

	
	\author{Fehmina Malik, Manjesh K.~Hanawal,
		and Yezekael Hayel 
		\IEEEcompsocitemizethanks{\IEEEcompsocthanksitem Fehmina Malik and Manjesh K. Hanawal are with IEOR, IIT Bombay, India. E-mail: \{fehminam, mhanawal, \}@iitb.ac.in. Yezekael Hayel is with LIA/CERI, University of Avignon, France. E-mail: yezekael.hayel@univ-avignon.fr. 	}
	}
	
\maketitle
	
	\begin{abstract}
It has been a long demand of Internet Service Providers (ISPs) that the Content Providers (CPs) share their profits for investments in network infrastructure. In this paper, we study profit sharing contracts between a CP with multiple ISPs. Each ISP commits to improving the Quality of Service (QoS) for the end-users through higher investments efforts. The CP agrees to share the profits due to the resulting higher demand for its content. We first model non-cooperative interaction between the CP and the ISPs as a two-stage Stackelberg game. CP is the leader that decides what fraction of its profits will be shared with the ISPs. Each ISP then simultaneously decides the amount of effort (investment) to enhance network quality. Here, CP cannot observe individual effort by the ISPs, which poses a challenge for the CP to decide how to share the profits with each ISP. Therefore, we also investigate a cooperative scenario, where the CP only decides the total share it gives to the ISPs, and each ISP then cooperatively shares the profit among themselves. We study the effect of such cooperation between the ISPs by building a Nash Bargaining based model. We show that the collaboration improves total effort by the ISPs and the payoff of the CP.
	\end{abstract}
	
	
	\section{Introduction}
In recent years, the growth of Internet traffic has been rapid, mainly due to an increase in the usage of data-intensive services like online video streaming, which accounts for over $65\%$ of the total downstream Internet traffic, of which Youtube contributes for over $15\%$ and Netflix over $11\%$ \cite{Sadvine1}.  Further, the COVID-19 pandemic led to the adoption of remote working, e-teaching, online collaboration, gaming, video streaming, etc., and almost $40\%$ growth in global traffic is seen within three months in early 2020 \cite{Sadvine2}. All these sudden changes have put unprecedented stress on the network, causing congestion. Internet service providers (ISPs) have to upgrade the network infrastructures to address the network performance issue in this fast-growing Internet traffic environment. Besides increasing network bandwidth/capacity, deployment of caching technologies is viewed as a critical solution for the upgradation of networks to improve the Quality of Service (QoS) for end-users \cite{leighton2009improving}. 

The QoS perceived by users is the primary concern for content providers (CPs). The high service quality directly increases users' desire for content and generates more demand for CPs, leading to higher revenue (mainly through advertisements and subscriptions). As network congestion causes poor content quality, resulting in decreased users' willingness to browse the content thus, CPs have incentives to pay for additional network investment to elevate the QoS for end-users. With ISPs' investment in the network, end users can enjoy a better QoS, which increases the demand and attracts more traffic, resulting in higher revenue of the CP.  Thus, ISP's investment in the network is beneficial to the CP and may share its profit with ISP to cover the investment cost. This increases network performance and, hence, covers the cost of the ISP's network investment. Such arrangements are becoming common between CPs and ISP, where ISP helps CPs to increase its demand by deploying caching resources or through premium peering (\cite{ bastug2014living, falkner2000overview}). For example, Netflix deploys a local cache of its most popular content within the data centers of partner ISP's network to enhance video quality for end-users (\cite{Netflix, Times, Comcast}).

Profit-sharing contracts between CPs and ISPs may be mutually beneficial to all and also benefit users. Through contracts, CPs may share their profits with ISPs to invest in network infrastructure. With the improved networks, users enjoy better quality over the Internet and increase demand for the content.  However, such private agreements may create disputes, raising policy concerns related to net-neutrality regulations. In this paper, we investigate the problem of profit-sharing between CP and multiple ISPs. In this case, the CP can only observe the overall impact on traffic, which depends on infrastructure enhancement (effort/action) by all ISPs, and cannot identify and distinguish individual efforts levels by each ISP. This poses a challenge for the CP to decide on profit sharing contract with each ISP.  We also consider the case of public ISP (owned by the government), which is non-profitable.
Further, we consider cooperative behavior where all players aim to maximize social utility. Such cooperative setups are known to benefit all players in \cite{mitra2019consortiums,malik2020revenue}). We consider a broad perspective on the behavior of the CP and ISPs to provide insights into new ways of dealing with profit-sharing between CP and ISPs.

In this paper, we consider different scenarios between ISPs. We first consider the case involving non-profitable public ISP and private ISPs, which are part of the marketplace. We then consider the case involving the private ISPs, but they can act cooperatively or non-cooperatively. Our motivation to study these different scenarios is to investigate which market framework is more beneficial for end-users and with a better QoS. Our contributions and observations are as follows:

\begin{itemize}
		\item We introduce Profit-sharing contracts between a CP with multiple ISPs. The CP first offers a contract to each ISP, and ISPs accordingly determine their optimal effort to improve demand for the CP's content. This results in a Stackelberg game with a single leader (CP) and multiple followers (ISPs). 
		\item We consider several market frameworks: (i) One public (non-profitable) and one private ISP that maximizes its utility, and (ii) Private ISPs in competition or cooperation.
		\item We show that if there is no obligation on CP to sign a contract with Public ISP, participation of public ISP has no effect. While if CP is obliged to sign the profit contract with Public ISP, it will end up sharing more profit with ISP while getting the same demand increment as in the case when CP is not obliged to make a deal with the public ISP.
		\item The equilibrium contracts is analyzed with private ISPs in non-cooperative and cooperative scenarios. In the non-cooperative scenario, CP can offer a customized profit share proportion for each ISP, and ISPs decide on their efforts to maximize their profits. In the cooperation scenario, a mechanism is proposed where CP makes a joint payment to ISPs where the ISPs bargain with each other to make decisions on their individual efforts such that profit share is in the proportion of their effort.
		\item When the ISPs are symmetric with respect to per unit cost for investment, we show both cooperative and non-cooperation scenarios results in the same level of efforts and demand.
		\item When the ISPs are asymmetric with respect to per unit cost for investment, the cooperative mechanism imposed over ISPs alleviates the utility of all ISPs and CP while also increasing the QoS for the content from CP.
		\end{itemize}

This paper is organized as follows. In Section \ref{problem} we discuss the problem setup and define the contracts. We study the
equilibrium contracts for one public and one private ISP case  in Section \ref{publi-privateisps}. We further studied competitive and cooperative scenario symmetric ISP case in Section \ref{symmetric}. The
asymmetric ISP case is studied in Section \ref{asymmetric} and the comparison between public and private ISPs (competition and cooperation) is discussed in Sections \ref{publicprivate}-\ref{compvscoop}. Conclusions are discussed in Section \ref{conclusion}. Proofs of all stated results can be found in the appendix.	
	\subsection{Related work:}
	There are a number of papers which explored the possibility of revenue sharing or surplus transfer between service providers in the internet (\cite{he2005pricing,saavedra2009bargaining,ma2010,ma2011, kamiyama2014effect,kamiyama2014feasibility,park2014isp1,park2014isp2,im2016revenue,yeze2017,kalvit2019capacity, mitra2019consortiums,badasyan2008simple, dhamdhere2010value, wang2021paid, kesidis13,kunsemoller2017game, garmani2018caching, shi2019optimal,shi2020multi}). Content charges by ISP to recover investment costs has been studied in \cite{he2005pricing,kamiyama2014effect, kamiyama2014feasibility,
park2014isp1, park2014isp2,im2016revenue}. In \cite{kamiyama2014effect} and \cite{kamiyama2014feasibility}, the authors investigated the feasibility of ISPs charging the CPs, and evaluated its effect by modeling the Stackelberg game between CPs and
ISPs. In \cite{park2014isp1} and \cite{park2014isp2}, a revenue-sharing scheme is proposed when the ISP provides a content piracy monitoring service to CPs for increasing the demand for their content. This work is extended to two ISPs competing with each other in \cite{im2016revenue} where only one of them provides the content piracy monitoring service. While some studies proposed cooperative settlement between service providers through Shapley value  \cite{ma2010,ma2011,yeze2017}) or through Bargaining concept (\cite{saavedra2009bargaining, kalvit2019capacity,mitra2019consortiums}). Some works (\cite{ badasyan2008simple, dhamdhere2010value, wang2021paid}) considered the problem of peering agreements to extract surplus from CP to ISP, where CP pays ISP to invest in peering link to improve QoS.  \cite{ shi2019optimal} discussed the problem of single CP with single non-strategic ISP, with both premium peering and caching, while \cite{shi2020multi} extended it for multiple CPs. They considered ISP to be non-strategic, and prices of resources are determined by market equilibrium. Some works consider the payment for the deployment of CDNs for in-network caching to enhance QoS (\cite{kesidis13,kunsemoller2017game, garmani2018caching}). 

Our previous work \cite{malik2020revenue} considered the contract design problem between monopolistic ISP and CPs, both in the presence and absence of neutrality regulations. In the present paper, we focus on a scenario with multiple ISPs and a single CP. The main problems of interest are: 
\begin{enumerate}
\item How is the profit share split between ISPs, as CP only observes the overall impact on the demand due to sum investment by all ISPs?
\item What cooperative mechanism between ISPs can help elevate the total investment and hence the QoS for end-users without degrading ISPs’ utilities?   
\end{enumerate}
We consider a class of linear contracts between a CP and multiple ISPs that leads to a tractable analysis. We study different types of interactions and compare the impact of the contract on utilities of the CP and ISPs and QoS of end-users under competitive and cooperative scenarios. We distinguish between public and private ISPs and study how their participation benefits the end users.  


\section{Problem Description} \label{problem}
\begin{figure}[!b]
\centering
\label{fig:model}
\includegraphics[scale=0.4]{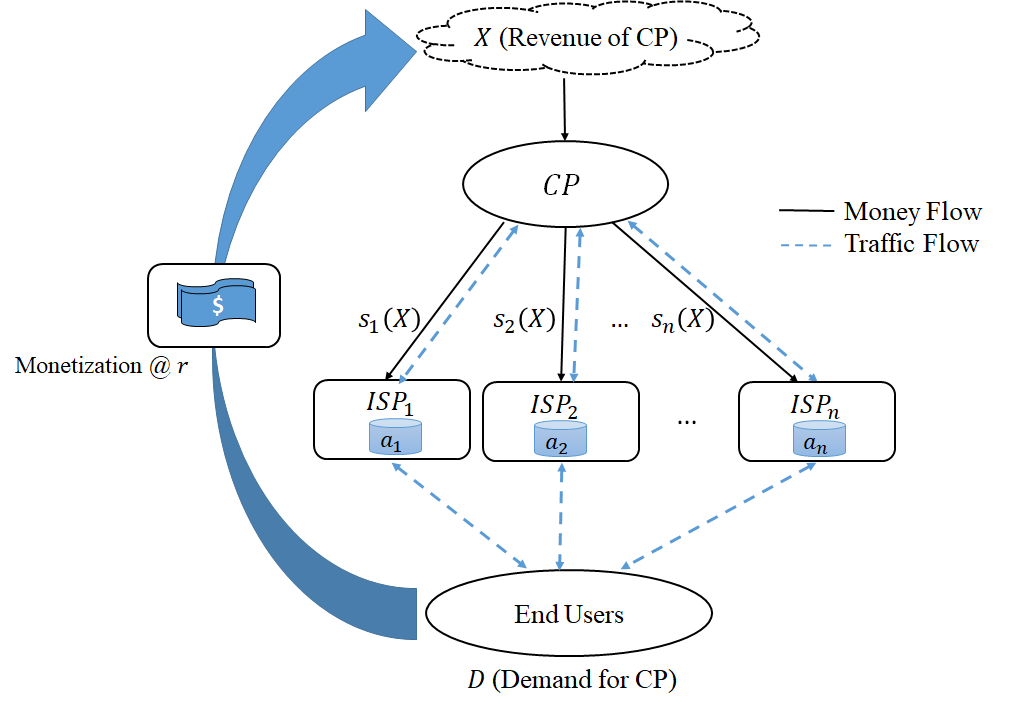}
\caption{Revenue flow between CPs, ISP and End-Users}
\end{figure}		
	
We consider a single Content Provider (CP) whose content can be accessed by multiple Internet Service Providers (ISPs). Let $n$ denote the number of ISPs and $\mathcal{N}=\{1,2,...,n\}$ denote the set of ISPs.
For each $i\in \mathcal{N}$ we denote the i-th ISP as $ISP_i$.  CP enters into an profit sharing contract with the ISPs in which $ISP_i, i \in  \mathcal{N}$ chooses to make some networks investment  quantified as $a_i \in \mathcal{R}_+;i \in \mathcal{N}$  to improve QoS for end-users. This investment, for example, could be related to caching efforts or improving the bandwidth of the network  The resulting increase/gain in the revenue of CP over a pre-specified horizon (say a billing cycle) is denoted by $X \in \mathcal{R}_+$ and depends on the total effort of ISPs. In return, CP shares part of its profits with ISPs to incentivize its investment; this share for each $ISP_i$ is determined by a sharing function $s_i: \mathcal{R}_+ \rightarrow \mathcal{R}_+$, which is also referred to as the contract/agreement. Specifically, the CP makes a payment of $s_i(X)$ to $ISP_i$ as part of this contract. Thus, the effective net revenue increase of the CP is given by the difference $ \left(X-\sum_{i \in \mathcal{N}}s_i(X)\right)$. Figure \ref{fig:model} represents the interactions between the agents in our model.

The increment in the demand for the content of CP denoted $D$ depends on the total efforts $\sum_{i\in \mathcal{N}}a_i$'s made by all $ISP_i$, $i \in \mathcal{N}$ jointly. Consistent with the law of diminishing returns, we model the increment $D$ as an increasing concave function of $a_i$'s \cite{shephard1974law}. For analytical tractability, we assume that $D$ grows logarithmically in $\sum_ia_i$'s and is given by $D := \log(1+\sum_ia_i)$. The resulting profit (increment) for CP is assumed to be proportional to the demand (increment) and is given by $X := rD$, where $r$ is a constant that captures how each additional unit of demand translates to earnings. Thus, the revenue generated by CP is then $X= r \log(1+\sum_ia_i)$. 
We restrict our analysis to linear revenue sharing contracts between ISPs and the CP. 
Specifically, these contracts are of the form $s_i(X) = \beta_i X$, where $\beta_i\in[0,1]$ for all $i \in \mathcal{N}$. Thus the contracts are paramterized by $\beta_i,i\in \mathcal{N}$. The utility of the CP is denoted as $U_{CP}$ and is given by
	$$U_{CP}=\left (1-\sum_{i \in \mathcal{N}}\beta_i\right )r\log\left ( \sum_{i \in \mathcal{N}}a_i+1\right).$$
	
Each ISP incurs a cost to improve the network quality. We assume that $ISP_i$ incurs a cost of $c_i$ per unit efforts and hence the total cost incurred by an effort level of $a_i$ is given by $c_ia_i$. Then the utility of $ISP_i$ is given by
	$$U_{ISP_i}=\beta_ir\log\left (\sum_{i \in \mathcal{N}}a_i+1\right )-c_ia_i.$$

We consider a leader-follower interaction as a Stackelberg game, with the CP acting as leader and ISPs as followers. The CP leads the system by announcing the revenue sharing contract, i.e., value $\beta_i$ to each $ISP_i, , i \in \mathcal{N}$. $ISP_i$ then responds to these contracts by determining its effort $a_i$. As the leader in the game, CP is assumed to be able to anticipate ISPs’ reactions to its action. Therefore, we analyze the game using the backward induction method and the equilibrium obtained is referred to as Stackelberg equilibrium.  Finally, note that followers are interacting through the demand increment $D$ and actions of each  impacts the utility function of all ISPs. Henceforth we refer to $D$ as simply demand.

\section{One public and one private ISP} \label{publi-privateisps}
In this section, we consider a special case with two ISPs, i.e., $n=2$ where one of it is a public ISP while the other is a private ISP. Public ISP is assumed to be nonprofit in nature and just aims to cover its marginal investment, for example ISP runs as a public (social) service managed by a government. On the other hand, private ISP is market oriented and chooses its investment decisions (effort) to maximize its utility. Without loss of generality, we suppose that $ISP_1$ is  public and $ISP_2$ is private. The hierarchical optimization  simplifies to the following optimization problem: 



\noindent
CP (Leader):
$$\max_{\beta_i\in[0,1], \sum_i \beta_i\le1} \quad (1-\beta_1-\beta_2)r\log(\overline{a}_1+a_2+1)$$
$ISP_1$ (follower: Public ISP)
$$\overline{a}_1 \quad \text{such that} \quad \beta_1r\log(\overline{a}_1+a_2+1)-c_1\overline{a}_1=0$$
$ISP_2$ (follower: Private ISP)
$$\max_{a_2 \ge 0} \quad \beta_2r\log(\overline{a}_1+a_2+1)-c_2a_2$$

\noindent
The total effort $\overline{a}_1$ by $ISP_1$ is such that the cost of the efforts is recovered from the profit shared by the CP, i.e., utility of $ISP_1$ is zero, while $ISP_2$ aims to put efforts that maximizes its own utility. Thus, the $ISP_1$ participates in the contract but only agrees to put efforts just enough to cover its expenses.
The optimal efforts by the ISPs in response to the contract
offered by CP, is given by the following proposition. 

\begin{proposition} \label{prop:effort-public,private}
For a given contract $(\beta_1,\beta_2)$ offered by CP,  optimal effort by each ISP is given by: 

$$
\overline{a}_1(\beta_1,\beta_2)=\frac{\beta_1 r }{c_1}\log\left(\frac{\beta_2 r}{c_2}\right)
$$
and
$$
a_2(\beta_1,\beta_2)=\frac{\beta_2 r}{c_2}-\frac{\beta_1 r }{c_1}\log\left(\frac{\beta_2 r}{c_2}\right)-1,
$$

\end{proposition}

\noindent
Aware of the response by the ISPs, the optimal strategy of CP is to offer a contract $(\beta_1^*,\beta_2^*)$ that maximizes its utility, i.e., solution of the following maximization problem:
$$\max_{\beta_i\in[0,1], \sum_i \beta_i\le1} \quad (1-\beta_1-\beta_2)r\log\left(\frac{\beta_2 r}{c_2}\right).$$
Note that the increment depends only on $ISP_2$ parameters (in particular $\frac{\beta_2 r}{c_2}$) because the other ISP is public and aims to have zero utility. Note that it is only interesting to consider the case $r > c_2$. Because, if $r<c_2$, total effort in this case $\beta r/c_2<1$ as $\beta\in (0,1)$, and it is not worthwhile for CP to make investments in order to increase the demand. Also, by definition, total effort becomes zero when $r<c_2$, thus $\beta_2$ does not impact the results discussed ahead. Henceforth, we assume for the rest of the section that $r>c_2$.

The following result characterizes the equilibrium contracts between CP
and each ISP. 
\begin{thm} \label{thm:beta-public,private}
The equilibrium contract between CP and each ISP is given by
\begin{equation}
   \beta_1^*=0 \mbox{  and  } \beta_2^*=\frac{1}{W\left(\frac{r }{c_2}e\right)}.
\end{equation}
\end{thm}
where $W(\cdot)$ is the LambertW function (see \cite{corless1996lambertw}). Since $W(\cdot)$ in strictly increasing and $W(e)=1$, it follows that $\beta_2^* \in(0,1)$ whenever $r/c_2>1$. Moreover, note that equilibrium fraction $\beta_2^*$ is a strictly decreasing function of the ratio $r/c_2$ as might
be expected: the ratio $r/c_2$ captures the relative monetization power of the CP per unit effort cost of the ISP. If CP's ability to monetize the efforts improves, then it needs lower efforts. 
Using Theorem 1, one can characterize the equilibrium effort for all the players of the market.
\begin{cor} \label{cor:utility-public,private}
\textcolor{black}{At equilibrium,}
\begin{itemize}
    \item Efforts by the ISPs are given by
\begin{equation}
   \overline{a}_1^*=0 \mbox{ and  } \quad a_2^*=\frac{ r}{c_2 W\left(\frac{r }{c_2}e\right)}-1.
\end{equation}
\item  Utility of the CP is given by
\begin{equation}
    U_{CP}=\frac{r\left[W\left(\frac{r }{c_2}e\right)-1\right]^2}{W\left(\frac{r }{c_2}e\right)}.
\end{equation}
\item Utility of ISPs is given by
\begin{align}
	U_{ISP_1}=0 \mbox{ and }	U_{ISP_2}&=r\left(1-\frac{2}{W\left(\frac{r}{c_2}e\right)}\right)+c_2
\end{align}
\end{itemize}
\end{cor}

Thus introducing a public ISP into the market appears to be a good idea to end users perspective, as it can result in improvement of the investments done by the private ISP. But in fact, when a CP makes a contract with a public and a private ISP, the public ISP does not get any share from the CP at equilibrium and hence the public ISP ends up not contributing anything to improving the quality of service for end users. Effectively, public ISP participation in the contract has no effect and problem reduces to a interaction between the CP and private ISP. 
In following section we consider several private ISPs that distribute the content traffic to end users.

If the regulator put restrictions that the CP has to make contract with Public ISP (i.e. $\beta_1>0$), CP will end up sharing more revenue, while getting same total demand. See Appendix \ref{app:positivebeta} for detail. 


\section{Symmetric Private ISPs} \label{symmetric}
In this section we consider the symmetric case where cost incurred by all the ISPs is the same, i.e., $c_1 = c_2=\ldots = c_n := c$. In other words,
the ISPs are symmetric with regards to the per unit cost incurred for investment in QoS for CP. In
this setting, we analyze the equilibrium contracts arising in the competitive as well
as cooperative setting, and the resulting surplus of the CP and the ISPs. 

\subsection{Competitive scenario}
In  this  competitive  scenario between  $n$  private  ISPs,   each  ISP  maximizes  its utility in order to  determine  its optimal  effort. The CP shares optimally its revenue  with  each  ISP  individually.  Here, the two level optimization problem is given as:\\
CP (Leader):\\
$$\max_{\beta_i\in[0,1], \sum_i \beta_i\le1} \quad (1-\sum_{i=1}^n\beta_i)r\log(\sum_{i=1}^na_i+1)$$
$ISP_i; i\in \mathcal{N}$ (follower: Private ISP)
$$\max_{a_i \ge 0} \quad \beta_ir\log(\sum_{i=1}^n a_i+1)-ca_i$$
By symmetry, it must be
such that $a_1=a_2=\cdots=a_n:=a$. Following proportion gives best response of each ISP to the contract offered by the CP.

\begin{proposition} \label{prop:symeffort-NC}
The optimal effort $a$ of each ISP in response to the contract offered by the CP, $\beta$ satisfies the following relationships:\\
$$na+1= \frac{n\beta_1 r}{c}=\frac{n\beta_2 r}{c}=\cdots=\frac{n\beta_n r}{c}.$$
\end{proposition}
Note that, because of symmetry between ISPs,  ISP's best response forces equal share proportion by the CP for all ISPs. Now, given ISPs' best response one can characterize the equilibrium contracts between the CP and each ISP as follows.
	\begin{thm} \label{thm:symbeta-NC}
	Let $r>c$. 
	The equilibrium contract between CP and $ISP_i, i\in \mathbf{N}$ are given by
		$$\beta_1=\beta_n=\cdots=\beta_n:=\beta=\frac{1}{n W\left(\frac{r}{c}e\right)}.$$
	\end{thm}
	For $r\le c$, $\beta=0$. This is because $W(\cdot)$ is strictly increasing and $W(e) = 1$. Moreover, total effort becomes $\beta r/c<1$, and it is not worthwhile for CP to make investment for growth of the demand, and by definition total effort becomes zero when $r<c$. Henceforth, we assume that $r>c$.
	Using Theorem 2, one can characterize the equilibrium effort of ISPs and utility of CP and ISPs as follows.
	
	\begin{cor} \label{cor:symutility-NC}
At equilibrium
		\begin{itemize}
	    \item The effort by each ISP is given by:
		$$ a=\frac{\beta r}{c}-\frac{1}{n}=\frac{r}{nW\left(\frac{r }{c}e\right)}-\frac{1}{n}$$
		\item The utility of CP is given by
		$$U_{CP}=(1-n\beta)r\log\left(\frac{\beta r}{c}\right)=r\frac{\left[W\left(\frac{r }{c}e\right)-1\right]^2}{W\left(\frac{r }{c}e\right)}$$
		\item The utility of each ISP is given by
		\begin{align*}
		   U_{ISP}&=\beta r \log\left(\frac{\beta r}{c}\right)-c\left(\frac{\beta r}{nc}-\frac{1}{n}\right)\\ &=r\left(1-\frac{n+1}{nW\left(\frac{r}{c}e\right)}\right)+\frac{c}{n}
		\end{align*}
	\end{itemize}
	\end{cor}
From the above corollary, it is evident that the utility of CP is independent of the number of ISPs, implying that in the case of symmetry between ISPs, CP doesn't care about how many ISPs enter the contract for infrastructure investment. However, the individual effort and utility of each ISP depend on the number of ISPs entering the contract with CP.

\subsection{Cooperative mechanism}
We now study the cooperative scenario,  where  ISPs cooperatively decide their effort, and CP jointly makes a single revenue-sharing contract with all ISPs.  We propose that the division of revenue share between ISPs depends on the ratio of their efforts. Having specified the CP’s behavior,  a transparent platform controlled by a regulatory authority facilitates the bargaining between ISPs to ensure that the cooperative mechanism is adequately enforced. Over here, ISPs can interact and bargain to arrive at individual effort to offer to the  CP.  We apply the  classical  Nash  Bargaining  Solution  (NBS)  to  capture  the outcome of this interaction \cite{saad2009coalitional}. Formally, the two levels optimization problem with constraints is described as follows:\\
	CP (Leader):
$$\max_{\beta\in[0,1]} \quad (1-\beta)r\log(\sum_{i=1}^n a_i+1)$$	
ISPs:
$$\max_{a_i\ge 0} \quad \prod_{i=1}^n\left(U_{ISP_i}(a_i)-d\right)$$
subjected to
$$\frac{\beta_i}{\beta_j}=\frac{a_i}{a_j} \quad \forall \;i\ne j,\mbox{ and } \sum_{i=1}^n \beta_i=\beta$$
where $U_{ISP_i}(a_i)=\beta_ir\log(\sum_{i=1}^n a_i+1)-ca_i$ is the utility of $ISP_i$ given the global CP's revenue share $\beta$. The disagreement point $d$
corresponds to the ISP utilities when they act non-cooperatively, i.e., the Nash Equilibrium between the ISPs. The last two constraints can be simplified to $\beta_i=\frac{\beta a_i}{\sum_{j=1}^n a_j} \quad \forall i=1,2,...,n$. Because of the symmetry, efforts and sharing are such that $a_1=a_2=...=a_n:=a$, and $\beta_1=\beta_2=...=\beta_n=\frac{\beta}{n}$. Thus, the two levels optimization problem with constraints can be reduced to:\\
CP (Leader):
$$\max_{\beta\in[0,1]} \quad (1-\beta)r\log(na+1)$$	
ISPs:
$$ \max_{a\ge 0} \quad \left(\frac{\beta }{n}r\log(na+1)-ca-d\right)^n$$
\begin{proposition} \label{prop:symeffort-C}
    The optimal effort by ISPs in response to the contract offered by the CP is given by
    \begin{equation}
        a(\beta)=\frac{1}{n}\left(\frac{\beta r}{c}-1\right).
    \end{equation}
\end{proposition}
Given ISPs' best response, we now characterize the equilibrium contract between CP and ISPs in the following Theorem. 
\begin{thm} \label{thm:symbeta-C}
The equilibrium contract between the CP and ISPs is given by
\begin{equation}
    \beta=\frac{1}{W\left(\frac{r}{c}e\right)}
\end{equation}
\end{thm}
Note that the total profit share given by CP in competitive scenario is same as that in the cooperative scenario. Further, from eqn (5), it can be easily seen that effort is also the same. Hence, the competitive and cooperative optimal solution coincides when ISPs are symmetric with respect to their cost. Therefore, when ISPs are symmetric, competitive scenario is as good as the cooperative scenario. Hence there is no need to force cooperation among ISPs with costly regulations as long of all of them have same sensitivity to cost of efforts.

	\subsection{The effect number of ISPs}
We now study the impact of number of ISPs (or scaling in $n$) on the efforts of ISPs' at equilibrium, and the surplus of each agent. Our main result is the following.
\begin{thm} \label{thm:ISPnumber}
Suppose that $r > c$, the non-zero equilibrium satisfies the following properties.
 \begin{enumerate}
     \item The fraction $\beta$ of revenue share each ISP gets is a strictly decreasing function of $n$, however, total share $n\beta$ remain unchanged with change in $n$.
     \item The effort by each ISP for CP is a strictly decreasing function
of $n$, even though the total effort is unaffected by the change in $n$.
     \item The utility of the CP does not depends on $n$.
      \item The utility of each ISP is strictly decreasing in $n$.
 \end{enumerate}
\end{thm}
Theorem 4 highlights that an increase in the number of ISPs does not affect CP. However, as more ISPs enter such profit sharing contract with the CP, contribution by the CP gets 'split' further between ISPs, thus each ISPs earning reduces.

\section{Asymmetric ISPs} \label{asymmetric}
In this section we consider the asymmetric case where each ISPs cost of investments is is not the same. This is captured by allowing $c_i\ne c_j$ for $i\ne j$. Our interest is to understand how disparity in the cost influences preference of
the players in the competitive and cooperative scenarios of the access market. For simplicity and clarity of the main insights, we focus on the case with two ISPs ($n = 2$) and without loss of generality, we assume that cost of $ISP_2$ is more than that of $ISP_1$, i.e., $c_2>c_1$.

\subsection{Competitive scenario}
We first consider the competitive scenario between two private ISPs, where each provider determine its effort to maximizes its utility. This effort impacts the CP revenue who shares part of it with each ISP individually. Here, the bi-level optimization problem is given as:\\
CP (Leader):\\
$$\max_{\beta_i\in[0,1], \sum_i \beta_i\le1} \quad (1-\beta_1-\beta_2)r\log(a_1+a_2+1)$$
$ISP_i; i=1,2$ (follower: Private ISP)
$$\max_{a_i \ge 0} \quad \beta_ir\log(a_1+a_2+1)-c_ia_i$$
The best response of each ISP to the contract (sharing value) offered by the CP can be obtained explicitly by solving the ISP optimization problem, given that the sharing coefficients $\beta_1$ and $\beta_2$ are fixed.
\begin{proposition} \label{prop:asymeffort-NC}
The optimal effort by ISPs in response to the contract offered by the CP is given by one of the three following cases:\\
\begin{itemize}
\item If $\frac{\beta_1}{\beta_2}=\frac{c_1}{c_2}$, then 
$$a_1,a_2>0\quad \& \quad a_1+a_2+1= \frac{\beta_1 r}{c_1}=\frac{\beta_2 r}{c_2}.$$
\item If there exists $\lambda\ge 0$ such that $\frac{\beta_1}{\beta_2}=\frac{c_1}{c_2+\lambda}$, then
$$a_1>0, a_2=0\quad \& \quad a_1+1=\frac{\beta_1 r}{c_1}=\frac{\beta_2 r}{c_2+\lambda }.$$
\item If there exists $\lambda\ge 0$ such that $\frac{\beta_1}{\beta_2}=\frac{c_1+\lambda}{c_2}$, then
$$a_1=0, a_2>0, \quad \& \quad  a_2+1=\frac{\beta_1 r}{c_1+\lambda}=\frac{\beta_2 r}{c_2 }.$$
\end{itemize}
\end{proposition}

ISPs' best response can then be plugged into the CP optimization problem to characterize the equilibrium contracts between the CP and each ISP. Note that the costs should satisfy the constraint $r>c_1+c_2$. Otherwise, total  effort becomes $\beta_1r/c_1 <1$, and it is not worthwhile for CP to make investment for growth of the demand. By definition total effort becomes zero when $r < c_1+c_2$.

	\begin{thm} \label{thm:asymbeta-NC}
	Considering that $r>c_1+c_2$, CP utility is maximized when  $\frac{\beta_1}{\beta_2}=\frac{c_1}{c_2}$  and the equilibrium contract between CP and $ISP_i$ is given by
		$$\beta_i^*=\frac{c_i}{(c_1+c_2)W\left(\frac{r}{c_1+c_2}e\right)} \mbox{ for all } i=1,2 $$
	\end{thm}
	Based on previous result in Theorem \ref{thm:asymbeta-NC}, the effort of each ISP at equilibrium satisfy the following relationship:
		$$  a_1^*+a_2^*=\frac{r}{(c_1+c_2)W\left(\frac{r }{c_1+c_2}e\right)}-1.$$
The change in strategy of one ISP will impact the strategy of the other ISP; and there is a continuum of best efforts of the ISPs but the overall effort is constant. There are many examples in literature with infinite number of equilibrium especially where the actions o f the players are coupled \cite{luo2019infinite}. There could be a unique responses of ISP if there is asynchronous decisions. For example, one ISP can be the historical operator and determines its effort first, before the decision of a new ISP. Therefore, the interaction structure between the ISPs corresponds to a Stackelberg framework, assuming that the historical ISP knows that a concurrent will come into the market after him.

It is interesting to note that if one of the ISPs decreases effort, it will still get the same revenue share from the CP as the total effort is same at equilibrium. As the competition results in continuum of equilibrium, some specific equilibrium points may be preferred by each of the ISPs which leads to instability. Also, such competitive scenario may lead to free riding by one of the ISPs, therefore, in the next section we propose cooperation between ISPs, whereby CP makes a joint payment to them and ISPs cooperatively decide the effort and the division of revenue share between them. 

	\subsection{Cooperative mechanism}
\begin{figure*}[t] 
	\centering
{\includegraphics[scale=0.45]{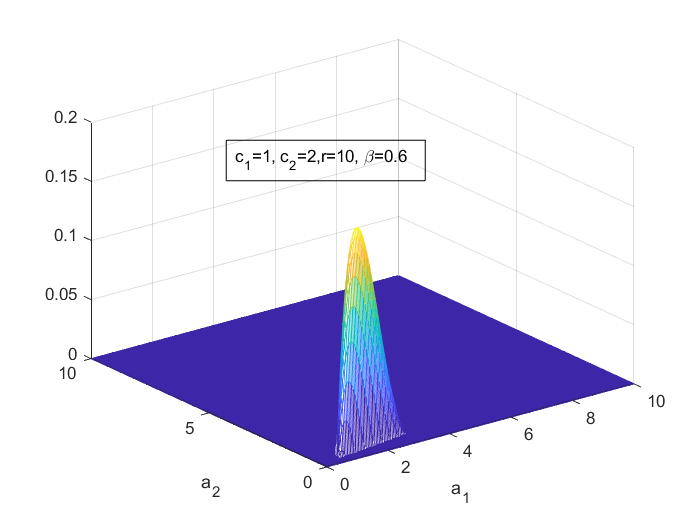}}
{\includegraphics[scale=0.45]{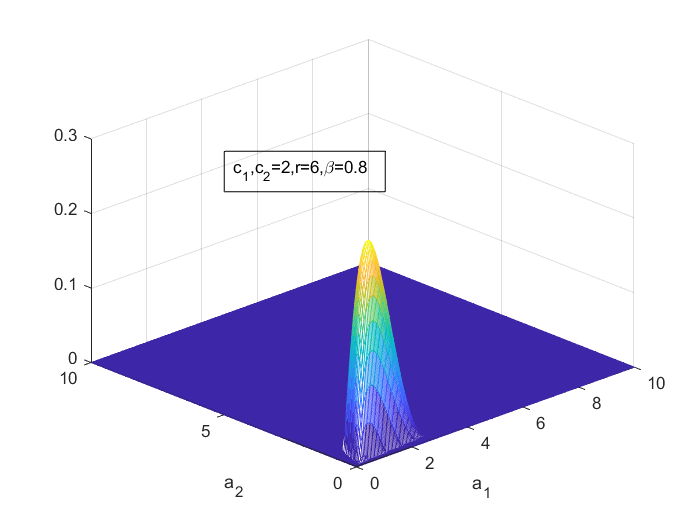}}
\caption{Graphical plots of objective function of NBP}
\end{figure*}
	We now study the cooperative scenario in which ISPs cooperatively decide their individual effort and CP makes a revenue sharing contract jointly with all ISPs. Formally, we define the bi-levels optimization problem as follows:\\
CP (Leader):
$$\max_{\beta\in[0,1]} \quad (1-\beta)r\log(a_1+a_2+1)$$	
ISPs:
$$\max_{a_1,a_2\ge 0} \quad \left(U_{ISP_1}(a_1,a_2)-d_1\right)\left(U_{ISP_2}(a_1,a_2)-d_2\right)$$
subjected to
$$\frac{\beta_1}{\beta_2}=\frac{a_1}{a_2}, \mbox{ and }\beta_1+\beta_2=\beta$$
where $U_{ISP_i}(a_1,a_2)=\beta_ir\log(a_1+a_2+1)-c_ia_i; i=1,2$ is the utility of $ISP_i$ given the CP's revenue share $\beta_i$. The disagreement point $d_i$ 
correspond to the $ISP_i$ utility in competitive scenario, i.e., the Nash equilibrium between the ISPs. Simplifying the last two constraints as $\beta_i=\frac{\beta a_i}{a_1+a_2} \quad \forall i=1,2$, the Bargaining problem for ISPs reduces to
\begin{align*}
   \max_{a_1,a_2\ge 0} \quad &\left(\frac{\beta a_1}{a_1+a_2}r\log(a_1+a_2+1)-c_1a_1-d_1\right)\times \\ 
   &\left(\frac{\beta a_2}{a_1+a_2}r\log(a_1+a_2+1)-c_2a_2-d_2\right)
\end{align*}

\begin{thm} \label{thm-NBS}
There exist a non-zero NBS, i.e. $a_1^B,a_2^B>0$. Furthermore, the non-zero solution $(a_1^B,a_2^B)$ is unique. 
\end{thm}
We also illustrate the concavity of objective function graphically in Figure 2. The graphs are three dimensional plots of the objective function with varying $a_1$ and $a_2$. The plots show unique peak, implying concave nature of objective function. 
The computation of NBS for this case is not analytically tractable because of the complex objective function arising due to asymmetrical costs of ISPs, but existence of non-zero solution and and uniqueness are established in the appendix. We the next section we empirically study how the cooperative mechanism affects all the players. The gradient methods can be used to numerically determine the NBS.

\section{Comparison between public/private ISP} \label{publicprivate}
We first compare the cases when CP makes contract with public and private ISPs verses both private ISPs. Our main result is the following.
\begin{thm} \label{thm:comparision}
The following statements hold with two competing ISPs:
\begin{enumerate}
    \item Total profit shared by the CP is higher when both ISPs are private as compared to the case where one ISP is public.
    \item Total effort is higher in the presence of public ISP as compared to the case when both ISPs are private.
     \item CP Utility is higher in the presence of public ISP as compared to the case when both ISPs are private.
\end{enumerate}
\end{thm}
The above Theorem implies that the introduction of public ISP makes private ISP invest more for the quality enhancement for end-users compared to the case of two private ISPs in competition. The higher investment results in higher demand for the CP, hence the higher utility for the CP. This means that competition between profit-maximizing ISPs skews the best effort by each ISP. Therefore, we proposed cooperation between ISPs  as an intervention and studied its impact in the next section. 

\section{Competition or Cooperation} \label{compvscoop}

\begin{figure*}[ht!] \label{fig:r}
	\centering
\subfloat[]{\includegraphics[scale=0.26]{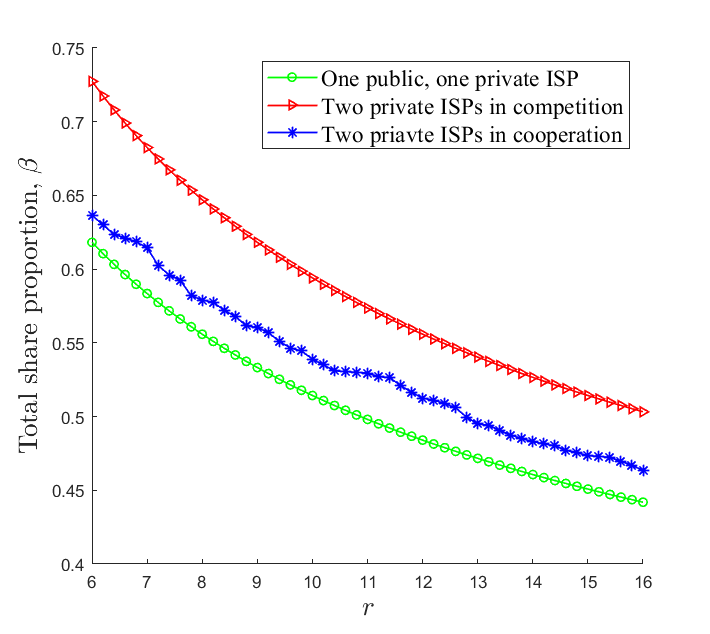}}
\subfloat[]	{\includegraphics[scale=0.26]{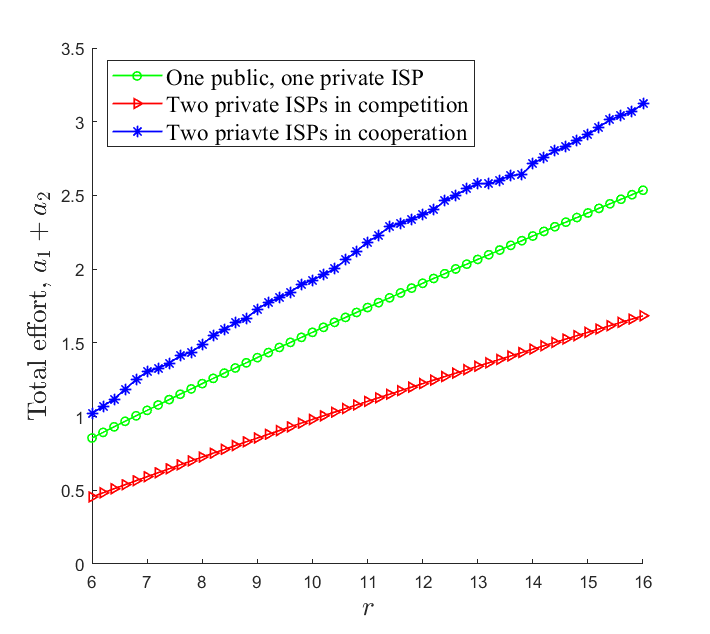}}
\subfloat[]	{\includegraphics[scale=0.26]{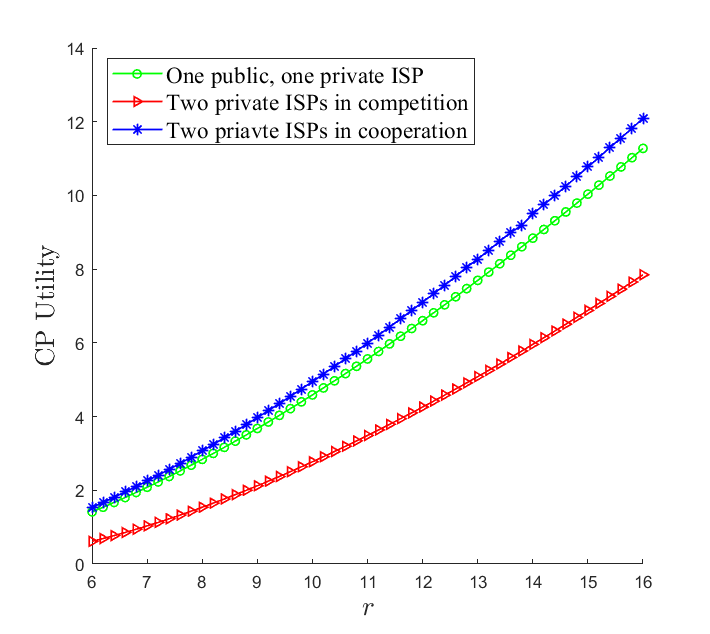}}

\subfloat[]	{\includegraphics[scale=0.26]{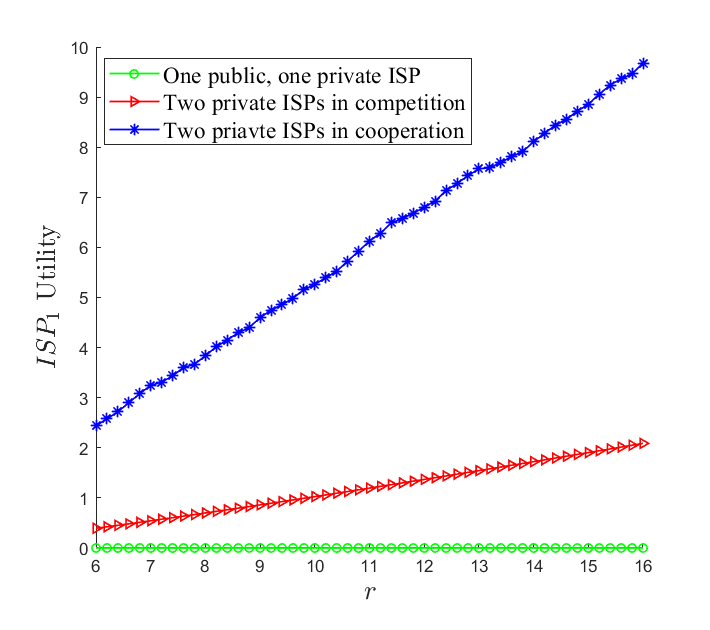}}
\subfloat[]	{\includegraphics[scale=0.26]{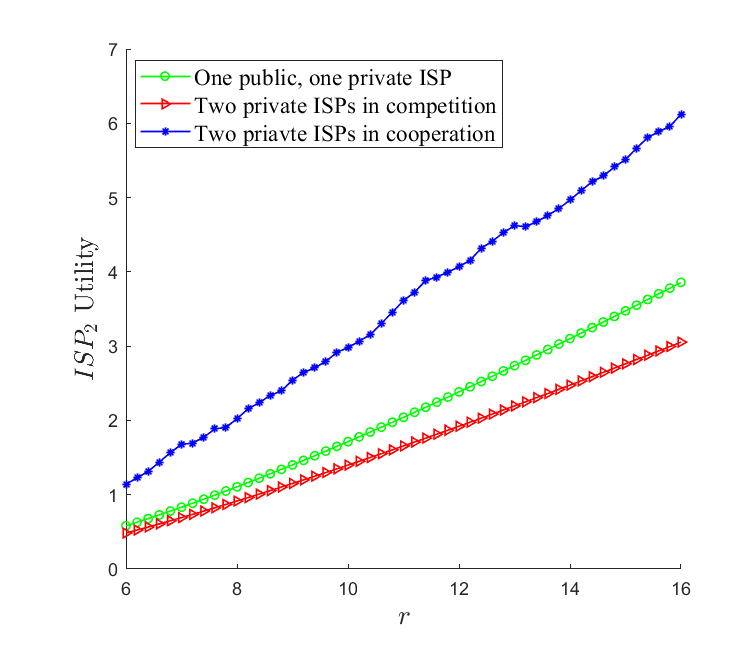}}
\caption{Comparison between three cases: One public ISP, two private ISPs in competition and cooperation. Here, we fix $c_2=2$, $c_1=1$ and vary $r$ from 6 to 12.}
\end{figure*}

\begin{figure*}[t!] \label{fig:c}
	\centering
\subfloat{\includegraphics[scale=0.26]{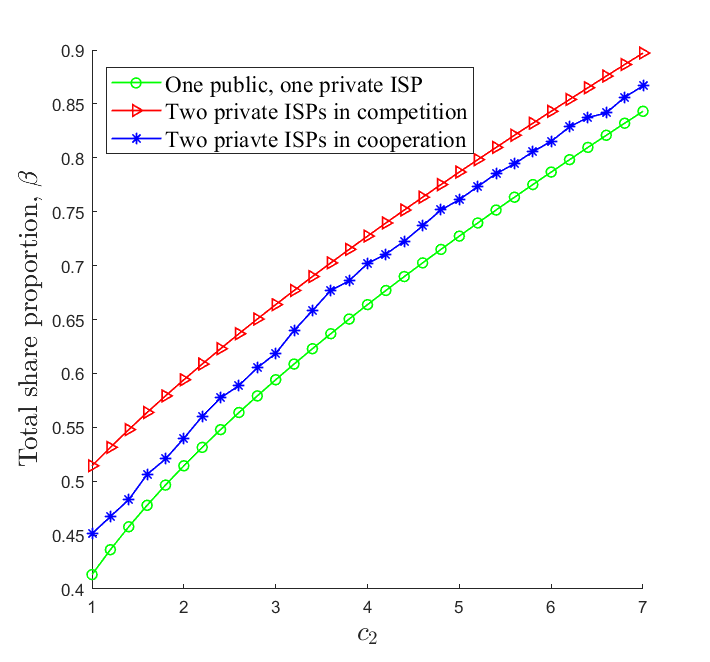}}
\subfloat	{\includegraphics[scale=0.26]{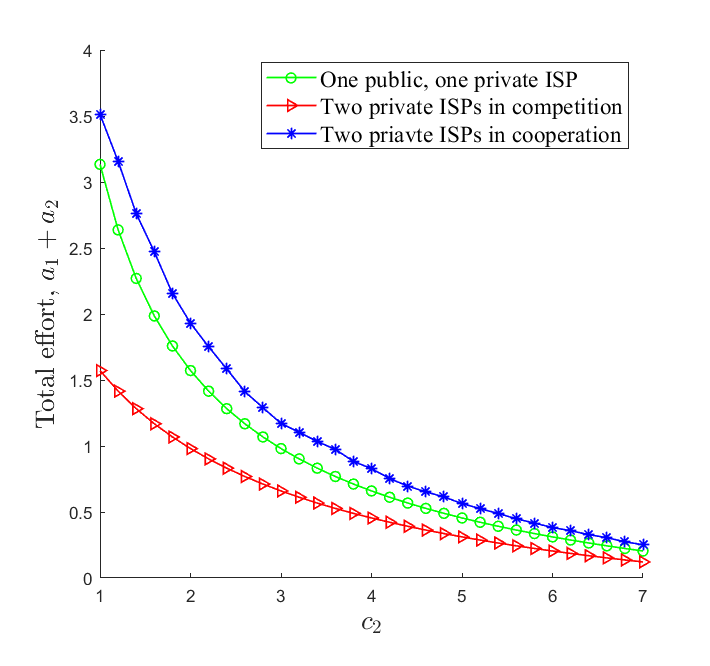}}
\subfloat	{\includegraphics[scale=0.26]{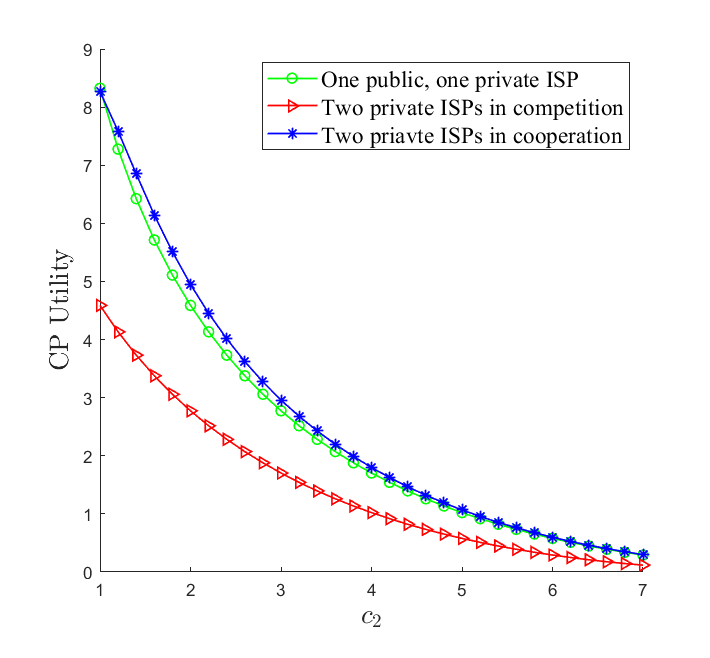}}

\subfloat{\includegraphics[scale=0.26]{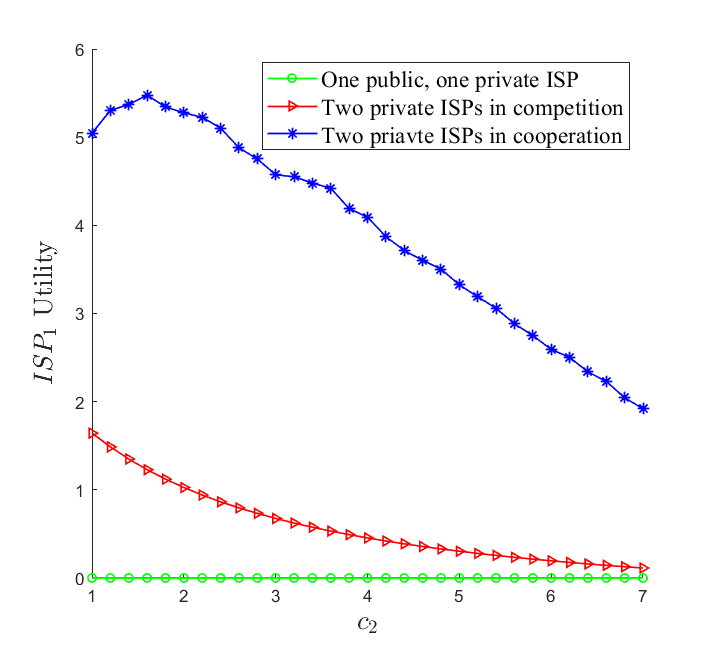}}
\subfloat	{\includegraphics[scale=0.26]{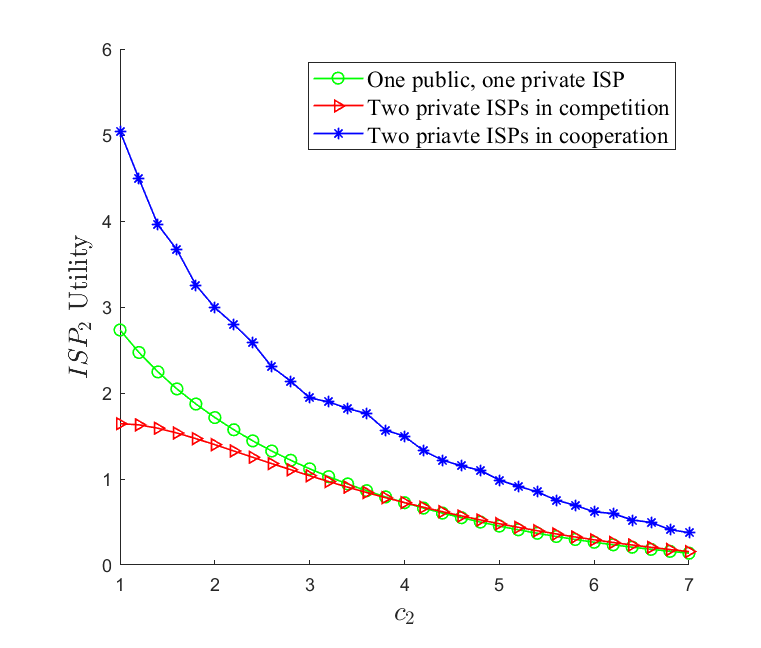}}
	\caption{Comparison between three cases: One public ISP, two private ISPs in competition and cooperation. Here, we fix $c_2=0.5=1$, $r=10$ and vary $c_2$ from 1 to 5.}
\end{figure*}

In this section, we compare the competitive and cooperative scenarios. We focus on the case of asymmetric ISPs, as in the section \ref{symmetric} we already show that for symmetric ISPs, both the scenarios give the same solution. Since analytical solution in cooperative solution is not tractable, we illustrate the impact of the proposed cooperative mechanism via numerical illustrations in Figures 3-4. Fig. 3(a)-(d) compares the three scenarios: public ISP, competition, and cooperation between private ISPs. We plot share proportion, total effort, and utilities of all agents by varying $r$, other parameters being fixed. Because of the infinite number of equilibrium between ISPs in competitive case, we choose the equilibrium for numerical illustrations such that $\frac{a_1^*}{a_2^*}=\frac{\beta_1^*}{\beta_2^*}$. As seen in figs. 3(b),(c)  that total effort and CP’s utility is highest when cooperative mechanism is imposed over ISPs. Note that the cooperative mechanism elevates the utility of both ISPs from the case when they act non-cooperatively. We also observe that share proportion is less in the cooperative case than in the competitive case and is least in the presence of public ISP. 

From Fig. 4(a)-(d), it is also clear that the share proportion is increasing in $c_2$, which is evident as the investment cost for ISP increases; the CP would have to contribute more revenue share to cover the cost of the ISPs. Further, the total effort by the ISPs decreases with $c_2$, implying higher the cost, the lesser the effort will be. The utility of all players decreases with an increase in $c_2$. The main reason for the decline is the decrease in total effort resulting in less demand, and hence CP earns less revenue resulting in less share from ISPs.

On the other hand, share proportion is decreasing in $r$, which implies that when the monetization power of CP is high, CP will have less incentive in revenue share contract for the investment cost of the ISP, as might be expected. The total effort by ISPs increases with $r$, as with the high monetization power of CP, ISPs have more incentive to invest in the network. Moreover, the utility of all players increases with an increase in $r$, which is because with higher monetization of CP, total effort increases resulting in higher demand and more revenue for the CP, hence the higher utilities for ISPs. 

As observed in the plots in Figures 3-4, it is evident that all the players are benefited in the cooperative mechanism, including the CP. Moreover, in all three scenarios, the total effort by ISPs is highest when ISPs cooperate, implying higher QoS for end-users. In fact, the cooperative mechanism elevates the utilities of ISPs and benefits CP and end-users. Therefore, everyone is better off by cooperation between ISPs.

\section{Conclusion} \label{conclusion}
We studied the problem of revenue sharing between single CP and multiple ISP on the Internet using the leader-follower interactions, with CP acting as a leader and the ISPs as followers. We consider the case where CP enters the profit sharing contract with both public (non-profitable) and private ISPs and show that having public ISP makes private ISP invest more in the network compared to the case of both being private ISPs.  Further, putting compulsion on CP to enter the contract with public ISP reduces CP’s surplus while having the same amount of QoS for end-users. We also studied the case where all ISPs are private, under competitive and cooperative setting. Cooperative mechanism is imposed over ISPs, where ISPs work together (like a coalition) and 'bargain' on a transparent platform to develop a mutually binding contract with the CP. We compared the total effort and utility of CP in these cases. We show that as expected, it is beneficial for everyone to have private ISPs in cooperation as it gives higher QoS for end-users and utility of each player as well. On the other hand, competition between private ISPs leads to less effort by ISPs which worsens the payoff for everyone.
	
	\bibliographystyle{IEEEtran}	
	\bibliography{bib}
	
	\section{Appendix}
\subsection{Proof of Proposition \ref{prop:effort-public,private}}
The optimization can be re-written as
\begin{align*}
 \max_{\beta_i\in[0,1], \sum_i \beta_i\le1} \quad &(1-\beta_1-\beta_2)r\log(\overline{a}_1+a_2+1)\\
\text{subjected to}&\\
 \overline{a}_1&=\frac{\beta_1r\log(\overline{a}_1+a_2+1)}{c_1}\\
 \max_{a_2 \ge 0} \quad \beta_2r&\log(\overline{a}_1+a_2+1)-c_2a_2
\end{align*}
Now, first order optimality condition for $ISP_2$ maximization problem is given by:
\begin{equation} \label{eqn:TAe}
    \frac{\beta_2 r}{\overline{a}_1+a_2+1}-c_2=0 \implies \overline{a}_1+a_2+1=\frac{\beta_2 r}{c_2}
\end{equation}
Using this, we get $\overline{a}_1=\frac{\beta_1r\log\left(\frac{\beta_2 r}{c_2}\right)}{c_1}$.
Substituting this $\overline{a}_1$ in eqn. (\ref{eqn:TAe}), we get
$
a_2=\frac{\beta_2 r}{c_2}-\frac{\beta_1 r }{c_1}\log\left(\frac{\beta_2 r}{c_2}\right)-1,
$

\subsection{Proof of Theorem \ref{thm:beta-public,private}}
Using proposition \ref{prop:effort-public,private}, the CP's optimization problem reduces to 
$$ \max_{\beta_i\in[0,1], \sum_i \beta_i\le 1} \quad \left(1-\beta_1-\beta_2\right)r\log\left(\frac{\beta_2 r}{c_2}\right)$$
It is optimized when $\beta_1=0$ and then CP's problem reduces to 
$$ \max_{\beta_2\in[0,1]} \quad \left(1-\beta_2\right)r\log\left(\frac{\beta_2 r}{c_2}\right)$$
 From CP optimization problem, it can be observed that for $r/c_2 < 1$, CP has no incentive to share a fraction of their revenue with the $ISP_2$ and $\beta_2=0$  is the equilibrium. Now assume $r/c_2 \geq 1$. For this case the optimal value of $\beta_2$ will be such that $r\beta_2/c_2\geq 1$. The first order optimality condition $\partial U_{CP}/\partial \beta_2 =0$ then gives:
$$\log\left(\frac{\beta_2r}{c}\right)=\frac{1-\beta_2}{\beta_2}$$
Solving and rearranging the terms, we get:
\begin{align*}
 \frac{1}{\beta_2}e^{{\frac{1}{\beta_2}}}&=\frac{re }{c_2}
\end{align*}
 Using the definition of the LamebertW function we get
$$\beta_2=\frac{1}{W\left(\frac{r}{c_2}e\right)}$$

\subsection{Proof of Corollary \ref{cor:utility-public,private}}
Substituting optimal $\beta_1,\beta_2$ obtained in Theorem \ref{thm:beta-public,private} in optimal effort, we get
\begin{align*}
a_1=0,	a_2&=\frac{r}{c_2W\left(\frac{r }{c_2}e\right)}-1
\end{align*}
\begin{align*}
    U_{CP}&=\left(1-0-\frac{1}{W\left(\frac{r }{c_2}e\right)}\right)\log\left(\frac{r}{c_2W\left(\frac{r }{c_2}e\right)}\right)\\
    &=\left(1-\frac{1}{W\left(\frac{r }{c_2}e\right)}\right)\log\left(\frac{\frac{re}{c_2}}{W\left(\frac{r e}{c_2}\right)e}\right)\\
    &=\frac{W\left(\frac{r }{c_2}e\right)-1}{W\left(\frac{r }{c_2}e\right)}\left(W\left(\frac{r }{c_2}e\right)-1\right) \\
    & \left(\because \log(\left(\frac{x}{W(x)}\right)=W(x)\right)\\
    &=\frac{\left[W\left(\frac{r }{c_2}e\right)-1\right]^2}{W\left(\frac{r }{c_2}e\right)} 
\end{align*}
$U_{ISP_1}=0$
\begin{align*}
    U_{ISP_2}&=\frac{r}{W\left(\frac{r }{c_2}e\right)}\log\left(\frac{\frac{r}{c_2}}{W\left(\frac{r }{c_2}e\right)}\right)-c_2\left(\frac{r}{c_2W\left(\frac{r }{c_2}e\right)}-1\right)\\
    &=\frac{r}{W\left(\frac{r }{c_2}e\right)}\left(W\left(\frac{r }{c_2}e\right)-1\right)-\frac{r}{W\left(\frac{r}{c_2}e\right)}+c_2\\
    &=r\left(1-\frac{2}{W\left(\frac{r}{c_2}e\right)}\right)+c_2
\end{align*}
\subsection{If regulator imposes that $\beta_1>0$}
\label{app:positivebeta}
We have the following problem:\\
\begin{align*}
 \max_{\beta_i\in[0,1], \sum_i \beta_i\le1} \quad &(1-\beta_1-\beta_2)r\log(\overline{a}_1+a_2+1)\\
\text{subjected to}&\\
 \overline{a}_1&=\frac{\beta_1r\log(\overline{a}_1+a_2+1)}{c_1}\\
 \max_{a_2 \ge 0} \quad \beta_2r&\log(\overline{a}_1+a_2+1)-c_2a_2
\end{align*}
Now, first order optimality condition for $ISP_2$ maximization problem is given by:
\begin{equation} \label{eqn:TA}
    \frac{\beta_2 r}{\overline{a}_1+a_2+1}-c_2=0 \implies \overline{a}_1+a_2+1=\frac{\beta_2 r}{c_2}
\end{equation}
Using this, we get 
\begin{equation} \label{eqn:a1}
\overline{a}_1=\frac{\beta_1r\log\left(\frac{\beta_2 r}{c_2}\right)}{c_1}    
\end{equation}
Then, the CP's optimization problem reduces to 
$$ \max_{\beta_i\in[0,1], \sum_i \beta_i\le 1} \quad \left(1-\beta_1-\beta_2\right)r\log\left(\frac{\beta_2 r}{c_2}\right)$$
Since, $\beta_1>0$, it will be set by the CP such that 
\begin{align} \label{eqn:beta1}
\overline{a}_1c_1={\beta_1r\log\left(\frac{\beta_2 r}{c_2}\right)}&\implies \beta_1=\frac{\overline{a}_1c_1}{r\log\left(\frac{\beta_2 r}{c_2}\right)}
\end{align}
Then CP's problem reduces to 
$$ \max_{\beta_2\in[0,1]} \quad \left(1-\beta_2\right)r\log\left(\frac{\beta_2 r}{c_2}\right)-c_1a_1$$
 From CP optimization problem, it can be observed that for $r/c_2 < 1$, CP has no incentive to share a fraction of their revenue with the $ISP_2$ and $\beta_2=0$  is the equilibrium. Now assume $r/c_2 \geq 1$. For this case the optimal value of $\beta_2$ will be such that $r\beta_2/c_2\geq 1$. The first order optimality condition $\partial U_{CP}/\partial \beta_2 =0$ then gives:
$$\log\left(\frac{\beta_2r}{c_2}\right)=\frac{1-\beta_2}{\beta_2}$$
Solving, we get:

$$\beta_2=\frac{1}{W\left(\frac{r}{c_2}e\right)}$$
Substituting this, we get
\begin{align*}
	a_2&=\frac{r}{c_2W\left(\frac{r }{c_2}e\right)}-\overline{a}_1-1\\
   \overline{a}_1&=\frac{\beta_1 r }{c_1} \left(W\left(\frac{r }{c_2}e\right)-1\right)
\end{align*}

 \subsection{Proof of Proposition \ref{prop:symeffort-NC}}
 Each $ISP_i$ optimizes over $a_i$ and by the symmetry we must have $a_1=a_2=...=a_n:=a$. Thus, the problem for $ISP_i$ is given as
 $$\max_{a\ge 0} \quad \beta_i r \log(na+1)-ca$$
 The first order necessary condition gives
 $$na+1=\frac{n\beta_i r}{c}\quad \forall i=1,2,...,n$$
 Then we must have 
 $$\frac{n\beta_1r}{c}=\frac{n\beta_2r}{c}=...=\frac{n\beta_nr}{c}\implies \beta_1=\beta_2=...=\beta_n$$
 \subsection{Proof of Theorem \ref{thm:symbeta-NC}}
 Given ISPs' best response, the problem for CP reduces to 
 $$\max_{\beta\in (0,1/n)} (1-n\beta)r\log\left(\frac{n\beta r}{c}\right)$$
 	The first order optimality condition $\partial U_{CP}/\partial \beta =0$ then gives:
$$\log\left(\frac{\beta r}{c}\right)=\frac{1-n\beta}{n\beta}$$
Solving and rearranging terms, we get:
$$\implies \frac{1}{n \beta }e^{{\frac{1}{n \beta}}}=\frac{re }{c}$$
Using the definition of the LamebertW function we get
 \[\frac{1}{n\beta}= W\left(\frac{re }{c}\right) \implies  \beta=\frac{1}{n W\left(\frac{r }{c}e\right)}\] 
 \subsection{Proof of Corollary \ref{cor:symutility-NC}}
 The proof is on the same lines as that of Corollary \ref{cor:utility-public,private}.
\subsection{Proof of Proposition \ref{prop:symeffort-C}}
First order condition for NBP between ISPs gives
$$\frac{\beta}{n}\frac{n}{na+1}-c=0 \implies na+1=\frac{\beta r}{c}$$

\subsection{Proof of Theorem \ref{thm:symbeta-C}}	
Given ISPs' best response, CP's problem reduces to
$$\max_{\beta\in [0,1]} (1-\beta )r\log \left(\frac{\beta r}{c} \right)$$
 
 	The first order optimality condition $\partial U_{CP}/\partial \beta =0$ then gives:
$$\log\left(\frac{\beta r}{c}\right)=\frac{1-\beta}{\beta}$$
Solving and rearranging terms, we get:
$$\implies \frac{1}{ \beta }e^{{\frac{1}{\beta}}}=\frac{re }{c}$$
Using the definition of the LamebertW function we get
 \[\frac{1}{\beta}= W\left(\frac{re }{c}\right)\implies \beta=\frac{1}{ W\left(\frac{r }{c}e\right)}\] 
 
\subsection{Proof of Theorem \ref{thm:ISPnumber}} 
\textit{Part 1:} We have
$$\beta(n)=\frac{1}{n W\left(\frac{re}{c}\right)}$$
It can be seen clearly that $\beta$ decreases with $n$, however, $n \beta=\frac{1}{ W\left(\frac{re}{c}\right)}$ remain unchanged with change in $n$.\\
\textit{Part 2:} The effort by each ISP is $$a(n)=\frac{1}{n}\left(\frac{r}{cW\left(\frac{re}{c}\right)}-1\right)$$
which is clearly decreasing function in $n$. 
The total effort $na=\frac{r}{cW\left(\frac{re}{c}\right)}-1$  remain unchanged with change in $n$.\\
\textit{Part 3:} Now, the utility of CP is given as
$$U_{CP}=(1-n\beta)r\log\left(\frac{n \beta r}{c}\right)$$
We know that $n \beta$ remain unchanged with change in $n$, therefore, from above expression it is clear that utility of CP also remains unchanged.\\ 
\textit{Part 4:} Sum utility of all ISPs is given by 
$$ nU_{ISP}=n\beta r \log\left(\frac{n \beta r}{c}\right)-c\left(\frac{n \beta r}{c}-1\right)
$$
Since, $n \beta $ remains same with change in $n$, from above expression it can be clearly seen that $nU_{ISP}$ also remains same with change in $n$. 
Hence, $U_{ISP}$ is strictly decreasing in $n$.

 \subsection{Proof of Proposition \ref{prop:asymeffort-NC}}
The problem for $ISP_i$ is given as
$$\max_{a_\ge 0}\quad \beta_i r \log(a_1+a_2+1)-c_1a_i$$
Lagrangian function for the problem is given as
$$L_i= \beta_i r \log(a_1+a_2+1)-c_ia_i-\lambda_ia_i$$
The corresponding KKT conditions are given as:
\begin{equation*} \label{eqn:KKT1}
    \frac{\partial L_i}{\partial a_i}=0 \implies a_1+a_2+1=\frac{\beta r}{c_i+\lambda_i}
\end{equation*}
\begin{equation*}\label{eqn:KKT3}
    \lambda_ia_i=0, \lambda_2a_2=0, \mbox{ and } \lambda_1,\lambda_2,a_1,a_2\ge0    .
\end{equation*}

Now, we have the following cases:
\textit{{Case 1:}} $a_1,a_2>0 \implies \lambda_1=\lambda_2=0$, then we get  
$$a_1=\frac{\beta_1r}{c_1}-a_2-1, \quad a_2=\frac{\beta_2r}{c_2}-a_1-1 $$
Solving, we get
$\frac{\beta_1}{\beta_2}=\frac{c_1}{c_2}$\\
\textit{{Case 2:}} $a_1=0, a_2>0 \implies  \lambda_1>0, \lambda_2=0$, then we have
$$a_2+1=\frac{\beta_1r}{c_1+\lambda_1}, \quad a_2+1=\frac{\beta_2r}{c_2}$$
Solving, we get
$\frac{\beta_1}{\beta_2}=\frac{c_1+\lambda_1}{c_2}\implies \frac{\beta_1}{\beta_2}>\frac{c_1}{c_2}$\\
\textit{{Case 3:}} $a_1>0, a_2=0 \implies  \lambda_1=0, \lambda_2>0$, then we have
$$a_1+1=\frac{\beta_1r}{c_1}, \quad a_1+1=\frac{\beta_2r}{c_2+\lambda_2}$$
Solving, we get
$\frac{\beta_1}{\beta_2}=\frac{c_1}{c_2+\lambda_2}\implies \frac{\beta_1}{\beta_2}<\frac{c_1}{c_2}$\\
\textit{{Case 4:}} $a_1=0, a_2=0 \implies  \lambda_1>0, \lambda_2>0$, This case is not possible because (0,0) cannot be equilibrium as it will give zero utility.
 \subsection{Proof of Theorem \ref{thm:asymbeta-NC}}
For Case 1: $\frac{\beta_1}{\beta_2}=\frac{c_1}{c_2}$, CP's problem is given as  
	$$
	\max_{\beta_i\in [0,1], \sum_i\beta_i\le 1} (1-\beta_1-\beta_2)r\log \left(\frac{\beta_1 r}{c_1}\right)
	$$
		$$
	\implies \max_{\beta_1\in [0,1]} \quad  \left(1-\left(\frac{c_1+c_2}{c_1}\right)\beta_1\right)r\log \left(\frac{\beta_1 r}{c_1}\right)
	$$
	The first order optimality condition $\partial U_{CP}/\partial \beta_1 =0$ then gives:
$$\log\left(\frac{\beta_1r}{c_1}\right)=\frac{1-\left(\frac{c_1+c_2}{c_1}\right)\beta_1}{\left(\frac{c_1+c_2}{c_1}\right)\beta_1}$$
Solving and using definition of LambertW function,we get:
 
 \[\beta_1=\frac{c_1}{(c_1+c_2)W\left(\frac{re }{c_1+c_2}\right)}\] 
 Using relation $\frac{\beta_1}{\beta_2}=\frac{c_1}{c_2}$, we get
	$$\beta_2=\frac{c_2}{(c_1+c_2)W\left(\frac{re}{c_1+c_2}\right)}$$
	$$U_{CP}^{I}=\left(1-\frac{1}{W\left(\frac{re}{c_1+c_2}\right)}\right)\log\left(\frac{\frac{r}{c_1+c_2}}{W\left(\frac{re}{c_1+c_2}\right)}\right)$$
Similarly for Case 2: $\frac{\beta_1}{\beta_2}=\frac{c_1+\lambda_1}{c_2}$, we get	
$$\beta_1=\frac{c_1+\lambda_1}{(c_1+c_2+\lambda)W\left(\frac{re}{c_1+c_2+\lambda_1}\right)}$$
$$\beta_2=\frac{c_2}{(c_1+c_2+\lambda_1)W\left(\frac{re}{c_1+c_2+\lambda_1}\right)}$$
$$U_{CP}^{II}=\left(1-\frac{1}{W\left(\frac{re}{c_1+c_2+\lambda_1}\right)}\right)\log\left(\frac{\frac{r}{c_1+c_2+\lambda_1}}{W\left(\frac{re}{c_1+c_2+\lambda_1}\right)}\right)$$
and for Case 3: $\frac{\beta_1}{\beta_2}=\frac{c_1}{c_2+\lambda_2}$, we get	
$$\beta_1=\frac{c_1}{(c_1+c_2+\lambda_2)W\left(\frac{re}{c_1+c_2+\lambda_2}\right)}$$
$$\beta_2=\frac{c_2+\lambda_2}{(c_1+c_2+\lambda_2)W\left(\frac{re}{c_1+c_2+\lambda_2}\right)}$$
$$U_{CP}^{III}=\left(1-\frac{1}{W\left(\frac{re}{c_1+c_2+\lambda_2}\right)}\right)\log\left(\frac{\frac{r}{c_1+c_2+\lambda_2}}{W\left(\frac{re}{c_1+c_2+\lambda_2}\right)}\right)$$
As $W(x)$ and $g(x):=\frac{x}{W(x)}$ is increasing in $x$ ($\because \frac{\partial g(x)}{\partial x}=\frac{1}{W(x)(1+W(x))}>0$), then we have
$$U_{CP}^I>U_{CP}^{II}, U_{CP}^I>U_{CP}^{III}$$
\subsection{Proof of Theorem \ref{thm-NBS}}	
\textbf{Existence:} For $\beta\in (0,1)$, we have
\begin{align*}
   \max_{a_1,a_2\ge 0} \quad &\left(\frac{\beta a_1}{a_1+a_2}r\log(a_1+a_2+1)-c_1a_1-d_1\right)\times \\ 
   &\left(\frac{\beta a_2}{a_1+a_2}r\log(a_1+a_2+1)-c_2a_2-d_2\right)
\end{align*}
$$\left(\frac{\beta a_1}{a_1+a_2}r\log(a_1+a_2+1)-c_1a_1-d_1\right)=F_1\text{(say)}$$
and
$$\left(\frac{\beta a_2}{a_1+a_2}r\log(a_1+a_2+1)-c_2a_2-d_2\right)=F_2\text{(say)}$$
Only possible solution to the NBP is $a_1, a_2>0$, because $a_1=0\implies F_1<0$ and $a_2=0\implies F_2<0$. 
Then, the NBS exists if and only if $F_1>0$ and $F_2>0$.\\
First consider $F_1>0$
\begin{align*}
   \iff & \frac{\beta a_1}{a_1+a_2}r\log(a_1+a_2+1)>d_1+c_1a_1\\
     \iff  & \frac{\beta }{a_1+a_2}r\log(a_1+a_2+1)>\frac{d_1}{a_1}+c_1 \quad (a_1>0)\\
   \iff & \beta>\frac{a_1+a_2}{\log(a_1+a_2+1)}\left(\frac{d_1}{a_1r}+\frac{c_1}{r}\right)
&\end{align*}
Since, $\beta\in(0,1)$, the above inequality holds iff
\begin{align} \nonumber
   \frac{a_1+a_2}{\log(a_1+a_2+1)}\left(\frac{d_1}{a_1r}+\frac{c_1}{r}\right)&<1\\\nonumber
   \iff \frac{d_1}{a_1r}+\frac{c_1}{r}<&\frac{\log(a_1+a_2+1)}{a_1+a_2}\\ \nonumber
     \iff \frac{c_1}{r}<&\frac{\log(a_1+a_2+1)}{a_1+a_2}-\frac{d_1}{a_1r}
\end{align}
Similarly, $F_2>0$
\begin{align} \nonumber
      \iff \frac{c_2}{r}&<\frac{\log(a_1+a_2+1)}{a_1+a_2}-\frac{d_2}{a_2r}
\end{align}
Since, $c_1<c_2$, we get
$$\frac{c_1}{r}<\frac{\log(a_1+a_2+1)}{a_1+a_2}-\frac{d_2}{a_2r}$$
Now, $\frac{\log(a_1+a_2+1)}{a_1+a_2}<1$, also $c_1/r<1 \forall i=1,2$. Therefore, from above inequality, it can be easily concluded that there will exist $a_1, a_2>0$ which satisfies this condition. \\
\textbf{Uniqueness:} Let $(\overline{a}_1,\overline{a}_2)$ and $(\hat{a}_1,\hat{a}_2)$ be two different Nash Bargaining solutions, i.e, two different optimizers of NBP. 
Then, we should have
$$F_1(\overline{a}_1,\overline{a}_2)F_2(\overline{a}_1,\overline{a}_2)=F_1(\hat{a}_1,\hat{a}_2)F_2(\hat{a}_1,\hat{a}_2)$$
Substituting the values and comparing the coefficient of terms on both the sides, we get
$$c_1d_2\overline{a}_1=c_1d_2\hat{a_1}\implies \overline{a}_1=\hat{a}_1$$
$$c_2d_1\overline{a}_2=c_2d_1\hat{a_2}\implies \overline{a}_2=\hat{a}_2$$

\subsection{Proof of Theorem \ref{thm:comparision}}
We use superscript $Pr$ and $Pu$ for the case where both ISPs are private and one ISP is public, respectively. \\
We have $\beta^{Pr}=\frac{1}{W\left(\frac{re}{c_1+c_2}\right)}$ and $\beta^{Pu}=\frac{1}{W\left(\frac{re}{c_2}\right)}$\\
Since, $W(x)$ is slowly increasing function in $x$, we have
\begin{equation}  
 {W\left(\frac{re}{c_1+c_2}\right)} <{W\left(\frac{re}{c_2}\right)}\implies \beta^{Pr}>\beta^{Pu}  \label{eqn:beta-comp}
\end{equation}
The total effort $TA^{Pr}=\frac{\frac{r}{c_2}}{W\left(\frac{re}{c_2}\right)}-1$ and $TA^{Pu}=\frac{\frac{r}{c_1+c_2}}{W\left(\frac{re}{c_1+c_2}\right)}-1$\\
Now, as $g(x):=\frac{x}{W(x)}$ is increasing in $x$, then we have 
\begin{equation} \label{eqn:TA-comp}
    TA^{Pu}>TA^{Pr}
\end{equation}
We can write the CP utility as 
$U_{CP}=(1-\beta)\log\left(TA+1\right)$
Now, using relations (\ref{eqn:beta-comp}) and (\ref{eqn:TA-comp}), we get
$$U_{CP}^{Pu}>U_{CP}^{Pr}$$

\end{document}